\documentclass[pra,twocolumn]{revtex4-1}
\usepackage[colorlinks,bookmarks=false,citecolor=blue,linkcolor=red,urlcolor=black]{hyperref}
\usepackage{graphicx}
\usepackage{amssymb,amsmath} 
\usepackage{amsfonts}
\usepackage{color}
\usepackage[utf8]{inputenc}
\usepackage[T1]{fontenc}
\usepackage{epsfig}
 
 \def\urlprefix{}
 \def\url#1{}
\def\be{\begin{equation}}
\def\ee{\end{equation}}
\def\bea{\begin{eqnarray}}
\def\eea{\end{eqnarray}}
\def\bi{\begin{itemize}}
\def\ei{\end{itemize}}
\def\bin{\begin{enumerate}}
\def\ein{\end{enumerate}}

\def\Nt{\tilde{N}}
\begin{document}
\title{Controlling disorder with periodically modulated interactions}
\author{Arkadiusz Kosior$^1$, Jan Major$^1$, Marcin P\l{}odzie\'n$^1$, and Jakub Zakrzewski$^{1,2}$} 
\affiliation{\mbox{$^1$Instytut Fizyki imienia Mariana Smoluchowskiego, Uniwersytet Jagiello\'nski, \L{}ojasiewicza 11, 30-348 Krak\'ow, Poland}
\mbox{$^2$Mark Kac Complex Systems Research Center, Jagiellonian University, \L{}ojasiewicza 11, 30-348  Krak\'ow, Poland}
}

\begin{abstract}
We investigate a celebrated problem of the one dimensional tight binding model in the presence of disorder leading to Anderson localization from a novel perspective. 
A binary disorder is assumed to be created by immobile, heavy particles that affect the motion of the lighter, mobile species in the limit of no interaction between mobile particles.
Fast, periodic modulations of interspecies interactions allow us to produce an effective model with  small diagonal and large off-diagonal disorder previously unexplored in cold atom experiments. We present an expression for an approximate Anderson localization length and verify the existence of the well known, extended resonant mode. We also analyze the influence of nonzero next-nearest neighbor hopping terms. We point out that periodic modulation of interaction allows disorder to work as a tunable band-pass filter for momenta.

\end{abstract}
\maketitle
\section{Introduction}
Anderson localization (AL) in disordered systems has fascinated and stimulated physicists for more than 50 years \cite{anderson58,Lagendijk2009}.
For one-dimensional (1D) systems, as discussed below, a particle propagating with  momentum $k$, in a disordered medium, undergoes multiple scatterings and eventually localizes with an exponentially decaying density profile \cite{mott61,lee85,ishii}. The AL is a single-particle interference effect which cannot be observed directly in solid-state systems due to the presence of electron-electron and electron-phonon interactions. AL has been widely studied for various systems including tight binding models with diagonal disorder and nearest-neighbor (nn)  and/or next-nearest-neighbor (nnn) hopping \cite{Sarma2011,Sepehrinia2010,Eilmes1998,Yamada2004}.

Ultracold atomic gases have become a playground where complex systems can be simulated and investigated \cite{Bloch2008, Jaksch2005,Lewenstein12}.  In particular, optical lattice engineering allows a high degree of controllability. Various techniques, i.e. periodic modulation of lattice positions and amplitudes  \cite{holthaus2005,Lignier2007,Struck2011} or Feshbach resonance \cite{Chin2010},  are used to effectively tune parameters of a given model \cite{sengstock2012,sengstock2013,targonska2012,eckardt2010,przysiezna2015}. The level of experimental control and detection allows one to build quantum simulators, i.e., experimentally controlled systems that are able to mimic other systems difficult to investigate directly \cite{quantum_simulators,Hauke2012}. Ultracold atomic gases are ideal systems for theoretical and experimental investigation of {the} Anderson localization of matter waves. An experimental observation of AL was first realized in one dimension seven years ago \cite{billy08} (a closely related 
Aubry-Andr\'e \cite{Aubry80} 
localization was also realized \cite{roati2008} independently. For reviews see \cite{Aspect09,Modugno10}. Recently  AL was observed  also in three dimensions \cite{anderson3D_jendrzejewski, 
anderson3D_kondov} in speckle potentials. In all these experiments, in order to get rid of particle interactions, either Feshbach resonances \cite{Chin2010} were employed or a low atomic density limit was reached. 

Typically in cold atom  disorder experiments, as well as theoretical propositions, the disorder appears  in a diagonal form, either on the chemical potential level (lattice site energies) like for  quasi-periodic lattices \cite{roati2008} or for interactions \cite{Gimp05}.  Similarly, a diagonal disorder appears for a binary disorder that is introduced by interactions with  other species \cite{gavish2003,Massignan06,Bongs_2006,horstmann2007,Graham2008}. 
On the other hand, a more detailed analysis shows that for cold atoms the off-diagonal disorder, i.e. a disorder in   tunnelings, appears in quite a natural way both for the incommensurate superlattice potential \cite{roscilde} and for the speckle random potential perturbing the optical lattice \cite{Hofstetter2010}.  However in both these cases, the disorder in tunneling is strongly correlated with the diagonal disorder. The aim of this paper is to show  
that periodic modulations  of interactions allow a transfer of the disorder to the kinetic energy (tunneling terms) creating a tunable
off-diagonal disorder opening up the possibility of its study in controllable cold atoms settings.

The interest in off-diagonal disorder stems from the fact that it can profoundly affect the properties of the system. Consider, for example, the case when the disorder is purely off-diagonal. There has been a long debate about the nature of states in the center of the band.
The first works \cite{bush75,Theodorou76,Antoniou77} showed that the localization length diverges in the center of the energy band  (i.e. $E=0$), therefore it was argued that a transparency window appears  and even the one-dimensional system exhibits the mobility edge. However, in the early eighties it has been established that even for a purely off-diagonal disorder all states are localized \cite{Soukoulis81,pendry1982,Delyon83,Krey83}.
The puzzle of the $E=0$ state was solved showing that the wavefunction envelope scales  as $exp(-\gamma\sqrt{N})$ with $N$ being the system size \cite{Soukoulis81}.  In the presence of both diagonal and off-diagonal disorder all states remain localized \cite{Bovier89} except very special cases of correlated diagonal and off-diagonal disorder \cite{flores1989}. Additional details, especially in the context of conductance anomalies, may be found in \cite{Alloatti09}. 

Possible correlations in the system and/or in the disorder may also profoundly affect localization properties \cite{DRDM,deMoura98,Piraud2013}. A famous example is the dimer model \cite{DRDM} in which sites may have energies 
$\epsilon_a$ and $\epsilon_b$ with the constrain that $\epsilon_b$-sites come only in pairs. This short-range correlation leads to delocalized states. A similar situation occurs for the dual  random dimer  model (DRDM) in which consecutive sites may not have  $\epsilon_b$ energy. The model has several applications in different areas from DNA studies \cite{Caetano,Ulloa2004} to photonic systems \cite{Zhao}. The cold atom version of DRDM has been proposed  by Schaff, Akdeniz, and Vignolo \cite{vignolo2010} who have shown that by tuning the interaction between atomic species one may observe the localization-delocalization transition and study the extended resonant mode. In our proposition we modify the approach proposed in \cite{vignolo2010} by periodically modulating the interactions which allows us a great freedom in changing the relative importance of diagonal and off-diagonal disorder in the system. For such a model we calculate the approximate Anderson localization length and compare it with numerical 
results. We show that the delocalization window may serve as a narrow energy filter for the particles.
  In addition we point out that the extended mode is vulnerable to effects due to next-nearest neighbor tunneling limiting its existence to relatively deep lattices.
Details of numerical methods needed to calculate the Anderson localization length with next-nearest neighbor random tunneling are presented in the Appendix B.

\section{The model \label{sec:PM}}
 
We consider a simple, standard one-dimensional non-interacting tight-binding Hamiltonian ({with} $\hbar=1$)
\begin{align}\label{Hm}
  H= \sum_i\left(\epsilon_i n_i-J( a^\dagger _i a_{i+1}+\mathrm{h.c.})\right),
\end{align}
where $a_i$ ($a^\dagger_i$) denotes an annihilation (creation) operator of bosons at site $i$, $n_i=a^\dagger_i a_i$ is a particle counting operator,  $\epsilon_i$ are on-site energies while $J$ is the tunneling amplitude which  will serve as our energy scale. Let us  apply a periodic modulation of the on-site energies 
\begin{align}\label{modula}
  \epsilon_i \to \epsilon_i(1+\delta\sin(\omega t)),
\end{align}
where $\delta$ is the modulation amplitude and $\omega$ its frequency.

Hamiltonian \eqref{Hm} has periodic time dependence $H(t)=H(t+2\pi/\omega)$. In such a case we can use the well established  formalism of Floquet theory \cite{Floquet} (see also \cite{Shirley}) extensively used, in the last century, to study the influence of optical or microwave fields  on atoms.  Recently it has been applied with great success for controlling properties of ultra-cold atomic systems \cite{holthaus2005,Lignier2007,Struck2011}. Solutions of equation $i\partial_t|\psi_n(t)\rangle=H(t)|\psi_n(t)\rangle$ have a form of Floquet states: $|\psi_n(t)\rangle=e^{-i\varepsilon_n t}|u_n(t)\rangle$, where $\varepsilon_n$ is called the quasienergy and $|u_n(t)\rangle$ have periodicity of the Hamiltonian. The Floquet theorem is a time analogue of the Bloch theorem for spatially periodic potentials. Although we can not treat $|\psi(t)\rangle$ as eigenstates of $H(t)$, $|u(t)\rangle$ are  eigenstates of the Floquet Hamiltonian $\mathcal{H}(t) = H(t)-i\partial_t$ existing in the extended space of $T$-periodic 
functions.
In that space we can number states using a new quantum number $m\in\mathbb{Z}$: $|u^m_n\rangle=|u_n\rangle e^{i\omega m t}$, where $|u_n^m\rangle$ is the eigenstate to the eigenenergy $\varepsilon^m_n=\varepsilon_n+\omega m$. This whole class corresponds to one physical state $|\psi_n(t)\rangle$ as adding $\omega$ to $\varepsilon_n$ in physical space only takes us to the next ``Brillouin zone''  for quasienergies.
Therefore, it is sufficient to find a block-diagonal form of a Hamiltonian (in $m$-ordered basis) and consider only one block corresponding to a single ``Brillouin zone''.
Unfortunately, couplings between different blocks make the block-diagonalization a formidable task. One may, however, make
 a unitary  transformation $U$:
\begin{align}\label{unitar}
\mathcal{H}'=U\mathcal{H}U^\dagger, \quad U=\exp{\left[-i \delta \frac{\cos(\omega t)}{ \omega}\sum_i \epsilon_i n_i\right]},
\end{align}
\be
\mathcal{H}'(t) =  \sum_i\left(\epsilon_i n_i-(e^{i(\epsilon_{i+1}-\epsilon_i)\frac{\delta \cos(\omega t)}{\omega}}a^\dagger _i a_{i+1}+\mathrm{h.c.})\right).
\ee
In effect, for frequencies $\omega\gg 1$ (in units of $J$) one we can neglect the couplings between different Floquet blocks and consider only one diagonal block governing the long term ($t \gg 1/\omega$) dynamics: 
\begin{align}\label{eq:H:PM1}
 H_{\mathrm{eff}}=\frac{1}{T}\int_0^{\frac{2\pi}{\omega}}dt\mathcal{H}'(t)=\sum_i\left(\epsilon_i n_i - (t_i a^\dagger _i a_{i+1}+\mathrm{h.c.})\right), 
\end{align}
where 
\begin{align} \label{tunel}
t_i=\mathcal{J}_0\left(\frac{\delta}{\omega}(\epsilon_{i+1}-\epsilon_{i})\right)
\end{align}
is the effective position dependent hopping and $\mathcal{J}_0$ is the zero-th order Bessel function. It has been verified experimentally \cite{Lignier2007,Struck2011} that the effective time-averaged Hamiltonian governs the dynamics of periodically driven systems for long times. In particular, localization properties of eigenstates of \eqref{eq:H:PM1} will be shared by the Floquet eigenstates of the original Hamiltonian \eqref{Hm} [with the fast modulation \eqref{modula}] after averaging them over the period.  {This may be understood by inspecting the unitary operator $U$ in \eqref{unitar} and observing that it consists of a product of local operators acting on single sites. Thus $U$ adds only local phases to single particle states -- that cannot affect the resulting density distribution.} Further detailed discussion of the construction of the effective Hamiltonians for high frequency periodically driven systems may be found in \cite{rahav03,eckardt15,bukov15}.   

Note that for a uniform system with  all $\epsilon_i$ being the same, the tunneling, \eqref{tunel} is unaffected, while $t_i$ changes if on-site energies vary from site to site. Such is the case  for superlattices \cite{esmann11,diliberto14} or external potentials such as a harmonic trap or a linear tilt \cite{holthaus2005}. In this work we will consider the on-site energy variations due to disorder. 

It is worth noting that it is usually possible to adiabatically pass from eigenstates of one $H_\mathrm{eff}$ to another (e.g. for different $\delta$) if the change of parameters is slow enough \cite{Poletti2011}. Thus it is possible to prepare a time independent system and turn on periodic modulation, or change modulation parameters during experiments. For completeness let us note that, for the tight-binding description in terms of a single lowest band to be valid in (\ref{Hm}-\ref{eq:H:PM1}), $\omega$, while much larger than the tunneling amplitude, must be smaller than the energy separation to the excited band \cite{holthaus2005,lacki2013}.

To create disorder we consider two species of atoms repulsively interacting with each other but allow only one of them to move freely on the lattice. The \emph{frozen} atoms (denoted with a superscript $f$) give rise to a binary disorder \cite{gavish2003,horstmann2007,horstmann2010,Bongs_2006}.  The mobile particles are assumed not to interact among themselves. They can be spin-polarized  fermions or bosons with interactions tuned off by microwave or optical Feshbach resonance \cite{Chin2010}. 
 The dynamics of the system can be described by a single particle Hamiltonian with the on-site energy
$\epsilon_i=V n^f_i,$ 
where $V$ denotes an interspecies contact interaction.
When \emph{frozen} particles are fermions or strongly repelling (hard-core) bosons their occupation per lattice site $n^f_i$ takes
either a value of zero or  one and the on-site energy takes only two values $\epsilon_i\in\{0,V\}$. Further, we consider the particular case  of DRDM when two adjacent sites cannot be occupied by \emph{frozen} atoms simultaneously \cite{DRDM}. For cold atoms DRDM  can be created  by the method described  in \cite{vignolo2010}. {Referring the reader to that paper for details, let us mention only, that to assure  no heavy particles reside in the neighboring sites one may first trap the heavy species in an auxiliary long-wavelength lattice with the lattice constant being e.g. three times bigger than the final lattice to be considered. If these heavy particles are strongly repelling and for sufficiently low densities one may assure no double occupancy.  Only then one can switch on the desired shorter wavelength lattice which holds the mobile component.}
 
In \cite{Rapp2012}  the authors considered fast periodic modulation of interactions induced by an appropriate  periodic modulation of the magnetic field $B(t) = B(t+T)$ with period $T = \frac{2\pi}{\omega}$ for interacting bosons in the optical lattice.
We assume a similar mechanism applied to light-heavy particle interactions resulting in periodic (for simplicity assumed to be harmonic) time dependence of site energies in the form $\epsilon_i(t)=n^f_i (V_0+V_1 \sin(\omega t))$,
where $V_1$ is an amplitude modulation strength and $\omega$ is the modulation frequency.  {For magnetic field values   close to the Feshbach resonance \cite{Chin2010}  large variations of interactions may be expected. In particular choosing the value of the mean magnetic field around which the oscillations occur one may vary at will the relative importance of the $V_0$ and $V_1$ coefficients. Note that we can adjust the magnetic field for this purpose since we assume that the mobile particles are either fermions or bosons with interactions turned off by
either microwave or optical Feshbach resonance \cite{Chin2010}. }

Compared to the general model 
discussed above we have $\epsilon_i= n^f_i V_0$ and $\delta=V_1/V_0.$ As $n^f_i\in\{0,1\}$, the effective tunneling, after time averaging, can only take two values:
 \begin{equation}
 \epsilon_i=\left\{\begin{matrix}
 0, & \mbox{ if } n^f_i = 0\\ 
 V_0, & \mbox{ if } n^f_i =1
\end{matrix}\right.,\quad
 t_i=\left\{\begin{matrix}
 1, & \mbox{ if } n^f_i = n^f_{i+1}\\ 
 1-\gamma, & \mbox{ if } n^f_i \neq n^f_{i+1}
\end{matrix}\right. ,
\end{equation}
where $\gamma=1-\mathcal{J}_0\left(V_1/\omega \right)$ measures the strength of off-diagonal disorder and varies in range from zero to about 1.4 (to the minimum of the Bessel function, $1+\min_x\mathcal{J}_0(x)$).

\section{Anderson localization length \label{sec:ALL}}
\begin{figure}
\includegraphics[width=7cm]{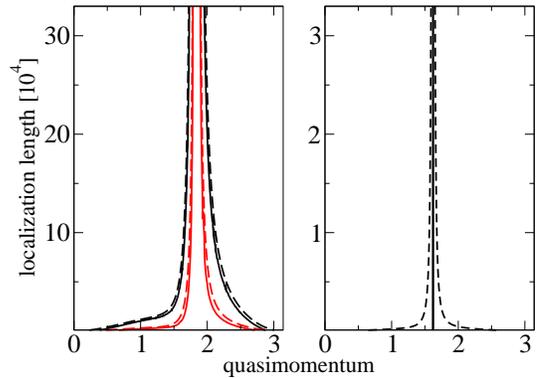}
\caption{(color online) 
Anderson localization length obtained from transfer matrix calculations (solid lines) and from (\ref{lambda}) (dashed lines).
Left panel: Anderson localization length as a function of quasimomentum for two sets of diagonal and off-diagonal disorder amplitude: $\{V_0=0.05,  \gamma=0.05\}$ (black) and $\{V_0=0.1,\gamma=0.1\}$ (red). Filling $\tilde{\rho}=1/3$. 
Right panel: Anderson localization length as a function of quasimomentum from transfer matrix calculation (black solid line) for diagonal disorder amplitude $V_0=0.1$, off-diagonal disorder amplitude $\gamma=0.9$ and occupation $\tilde{\rho}=1/3$. For comparison with (\ref{lambda}) we present analytical expression (black dashed line).}
\label{fig:loc_len}
\end{figure}

\begin{figure}
\includegraphics[width=6.5cm]{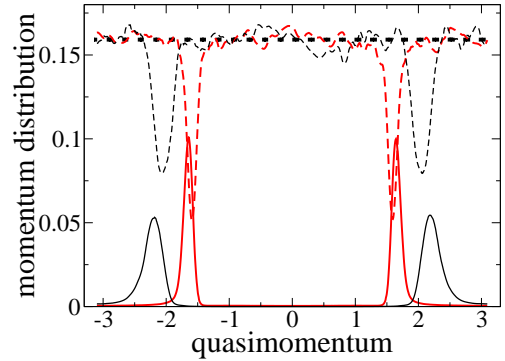}
\caption{(Color online)
Horizontal dotted thick line represents the initial wave packet momentum distribution. Solid lines present momentum distribution of wave function outside area of the disorder while dashed line the distribution of momenta that remain in the system.  $N = 1000$ lattice sites are considered with filling $\tilde{\rho} = 1/3$. Two cases are plotted. The data for diagonal and off-diagonal amplitudes $V_0 = 0.1$, $\gamma = 0.5$ are represented by red (thick) curves while those for $V_0 = 0.5, \gamma = 0.3$ by black (thin) lines. The evolution time is $t=1000$.
The position of dips in momentum space agree with position of transparent mode $k_t \approx 1.64$ and $k_t \approx 2.08$ respectively. One can observe two peaks in the momentum distributions of escaped atoms. Due to a marginal difference in the dispersion relation between in and outside of the disorder, the positions of peaks and dips are slightly shifted. These results are obtained as the average over $1000$ disorder realizations. }
\label{fig:laser}
\end{figure}

To calculate Anderson localization length let us start with the time-independent Schr\"{o}dinger equation for the disordered tight binding Hamiltonian \eqref{eq:H:PM1}
\begin{align}\label{eq:F1}
 -t_{i}\psi_{i+1}-t_{i-1}\psi_{i-1}+\epsilon_i \psi_i = \tilde{E} \psi_i,
\end{align}
where $\tilde{E}$ is the eigenenergy. In the regime of small diagonal and off-diagonal disorder ($V_0 \ll 1, \gamma \ll 1$)  we can assume
that $\tilde{E}$ is approximately given by the dispersion relation $\tilde{E} \approx-2\cos(k)$, where $k$ is the quasimomentum.
In order to define Anderson localization length in a system with off-diagonal disorder we utilize a unitary transformation 
$\psi_i=\phi_i \eta_i$, where $\eta_i=1/[t_i \eta_{i-1}]$. We transform equation~(\ref{eq:F1}) to diagonal form:
\begin{align}\label{eq:F:2}
 \phi_{i+1}+\phi_{i-1}+ \tilde{V_i} \phi_i = 0,
\end{align}
where $\tilde{V_i} = |\eta_i|^2(\epsilon_i+2\cos(k))$ is a new effective diagonal disorder. In the considered DRDM model  $\eta_i=1/t_i$.
 Equation (\ref{eq:F:2}) can be expressed as a two dimensional Hamiltonian map with position and momentum of the form $x_{i} = \phi_i$, $ p_{i}  = (\phi_i\cos(k) + \phi_{i-1})/\sin(k)$:
 \begin{align}\label{eq:map}
 \left\{\begin{matrix}
 x_{i+1} & = & -(p_i+A_i x_i)\sin(k) + x_i \cos(k)\\
 p_{i+1} & = & (p_i+A_i x_i)\cos(k) + x_i \sin(k),
 \end{matrix}\right.
 \end{align}
 where 
 \begin{align}
  A_i = -\frac{\epsilon_i-2(1-|t_i|^2)\cos(k)}{\sin (k)}.
 \end{align} 
The  map (\ref{eq:map}) describes the behavior of a harmonic oscillator under periodic delta kicks with amplitude $A_i$. Expressing
the map in action-angle variables and iterating it one may estimate the localization length in the limit of small diagonal and off-diagonal disorder. The details are given  in the Appendix.
We obtain the  approximate inverse localization length 
\begin{eqnarray}\label{lambda}
 \lambda^{-1}& = & \frac{\rho}{(1+\rho)^2}\frac{(V_0+2\gamma(2-\gamma)\cos (k))^2}{8\sin^2(k)}  \\
   & \times \nonumber & \left(1-2\frac{\rho(\rho+\cos(2k))}{1+\rho^2+2\rho\cos(2k)}\right),
\end{eqnarray}
where $\rho=\tilde{\rho}/(1-\tilde{\rho})$ and $\tilde{\rho}$ is the mean occupation number of \emph{frozen} particles.
A simple analysis of equation (\ref{lambda}) indicates an anomalous behavior of Anderson localization length due to the existence of a transparent mode with momentum $k_t$ given by condition $\cos(k_t) = -\frac{V_0}{2\gamma(2-\gamma)}$ for which the localization length diverges. 
\emph{Frozen} particles in a lattice increase the on-site energy and a motion of mobile particles can be seen as a sequence of scattering processes on potential barriers. In the DRDM  two \emph{frozen} particles are separated by at least one lattice site and the resonance is caused by a lossless transmission through a single barrier. The existence of the transparent mode is given by the condition $V_0\leq 2\gamma(2-\gamma)$ and the wave with momentum $k_t$ travels through the sample without reflection. A similar anomalous mode was observed in Ref.~\cite{vignolo2010}. The on-site energy modulation allows us to change the off-diagonal disorder significantly while keeping the amplitude of the diagonal disorder small 
(contrary to models deriving the changes of the tunneling from changes of an effective lattice shape only \cite{vignolo2010,Stasinska14}).
In Fig.~\ref{fig:loc_len} we present the numerically calculated Anderson localization lengths (using the standard transfer matrix method, see e.g.  \cite{delande2011}) and compare them with the analytical expression (\ref{lambda}). 
The left panel corresponds to the weak diagonal ($V_0 \in \{0.05,0.1\}$) and weak diagonal and off-diagonal ($\gamma\in \{0.05,0.1\}$) regime. The occupation of \emph{frozen} particles is equal to $\tilde{\rho} = 1/3$. The theoretical predictions are in good agreement with the numerical calculations. 
The right panel corresponds to the weak diagonal ($V_0 = 0.1$) and strong off-diagonal ($\gamma = 0.9$) regime. The width of the divergent Anderson localization length window is significantly narrower than for the weak disorder case. The position of the transparent mode $k_t$  is properly described by Eq. (\ref{lambda}) while the Anderson localization length shape is not.
Due to the  existence of the transparent mode in the system, one can expect that the evolution of an initial wave packet with a given momentum distribution results in Anderson localization of all momenta except those very close to the transparent mode $k_t$. Disorder effectively
works as a band-pass filter for momenta. To verify this behavior we integrated the Schr\"{o}dinger equation in time  for the Hamiltonian (\ref{eq:H:PM1}) with the initial state consisting of one particle in the center of the lattice with uniform momentum distribution. 
Fig.~\ref{fig:laser} presents momentum distributions of Anderson localized atoms and those that escaped from the disorder after time evolution $t = 1000$. The momenta of escaped atoms reveal narrow peaks at the position of the transparent mode while the momentum distribution of Anderson localized atoms reveal dips, positions of which agree with the former peaks. The lattice has $N = 1024$ sites with filling $\tilde{\rho} = 1/3$. We choose two sets of parameters for the amplitude of diagonal and off-diagonal disorder $(V_0 = 0.1, \gamma = 0.5)$ and $(V_0 = 0.5, \gamma = 0.3)$.  
For such parameters the transparent mode appears for  $k_t \approx 1.64$
and $k_t \approx 2.08$ respectively. 
Our system presents, therefore, a promising candidate for  
obtaining a highly controllable monochromatic gun for mater waves. A similar mechanism was observed in \cite{plodzien2011} where coherent multiple scattering processes determined the emitted matter-wave mode. In the following we shall inspect 
residual effects that may affect the existence of the transparent mode for a realistic system.  In particular, in the next Section we analyze  the  influence of the next-nearest neighbor tunneling on Anderson localization length in the regime of strong off-diagonal disorder.

\begin{figure}[h]
\begin{center}
\resizebox{1\columnwidth}{!}{\includegraphics{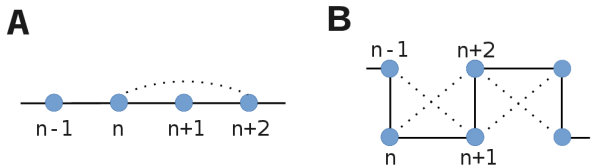}}
\caption{
Panel (a): One dimensional gas on a lattice with nearest neighbor hopping (solid line) and next-nearest neighbor hopping (dotted line). Panel (b): The same system can be viewed and described differently, namely as a stripe with non uniform tunneling. The advantage is that the hopping of particles takes place between neighboring slices only. 
}
\label{stripe}
\end{center}
\end{figure}

\begin{figure}[hbt]
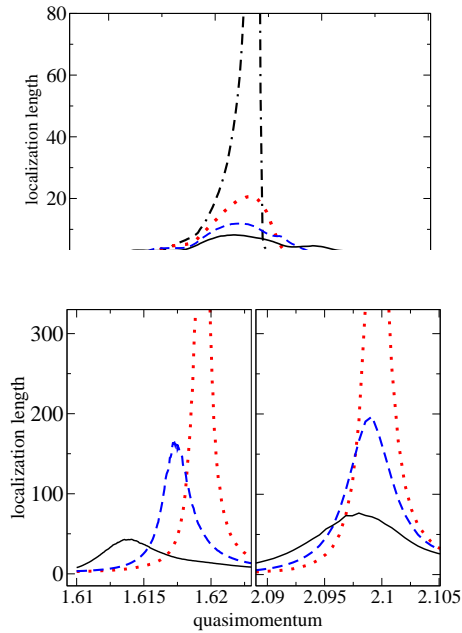

\begin{center}
\includegraphics[width=5.6cm]{zanik1c}
\includegraphics[width=6cm]{zanik23}
\caption{
(color online) Anderson localization length  as a function of quasimomentum for diagonal disorder $V_0=1.95$ (top panel), $V_0=1$ (right bottom) and  $V_0=0.1$ (left bottom),  $\gamma=0.9$, occupation $\tilde{\rho}=1/3$. The value of the next nearest neighbor hopping is $t'=0.01$ (black solid line), $t'=0.005$ (blue dashed line), $t'=0.0025$ (red dotted line)  and $t'=0.0005$ (black dashed-dotted line). The resonance diminishes when $t'$ grows. The effect is the strongest for $V_0=1.95$ because we are close to a regime without a delocalized mode.
}
\label{zanik_rezonansu}
\end{center}
\end{figure}
\section{Next-nearest neighbor tunnelings}

For lattice depths $U \approx 10$-$20\, E_R$, where $E_R$ is the lattice recoil energy and the typical optical laser wavelengths, next-nearest neighbor hopping amplitude $t'$ ranges between $0.01$-$0.001$, as can easily be estimated with appropriate Wannier functions. This is why in typical situations long range hopping is often negligible. Still let us add the next-nearest neighbor tunneling term to the Hamiltonian \eqref{Hm}
\be\label{add}
H\rightarrow H +t' \sum_i  (a^\dagger _i a_{i+2} +h.c.)\,.
\ee
Under the same unitary transformation (\ref{unitar}) and after time averaging we obtain 
the  effective Hamiltonian in the form of \eqref{eq:H:PM1} plus the additional term reading
\be\label{add2}
\sum_i t'_i(a^\dagger _i a_{i+2} +h.c.)\,
\ee
with
 $t'_i=t'\mathcal{J}_0\left(\frac{V_0}{\omega}(\epsilon_{i+2}-\epsilon_i)\right)$. 
   We numerically calculate Anderson localization length for this new effective Hamiltonian. The method we use is applicable to finite but arbitrarily long-range hopping.
 The transfer matrix approach is not directly applicable so we  transform the problem from a one dimensional to a two dimensional stripe.  This could be achieved with a simple winding of a string, see~Fig.~\ref{stripe}. In a stripe all particle hoppings are between the neighboring slices, which simplifies numerical calculations.

In the Appendix we present the derivation of the equation for a  Green's function between the first and the last slice of the stripe. Knowing the Green's function we can easily calculate the localization length \cite{osedelec, kramer1983}.

 As exemplary parameters for numerical analysis we choose $V_0=\{1.95, 1, 0.1 \}$ and $\gamma=0.9$  in order to obtain a regime 
 where a next-nearest-neighbor hopping significantly affects the dynamics of the system. The results are presented in Fig.~\ref{zanik_rezonansu}.  We observe that even extremely small values of $t'$ significantly affect the resonance. 
The presence of a small but non-zero next-nearest-neighbor tunneling $t'$ significantly reduces the  localization length for the transparent mode $k_t$ due to a non-zero probability that a mobile particle scatters back on a \emph{frozen} particle through a next-nearest-neighbor hopping. This process can appear when two \emph{frozen} particles are separated by a single site. Indeed, when we exclude such configurations, the resonance reappears. In deep lattices next-nearest-neighbor tunneling also  reduces the localization length for the transparent mode, but the effect is weaker and may not affect the selective emission due to a finite optical lattice size (when this size is smaller than the localization length).

\section{Conclusions \label{sec:CC}}
In this paper we proposed a method to realize a controllable off-diagonal disorder  with binary random potential resulting from time-periodic modulation of interactions between mobile and frozen particles in an optical lattice.
 Since no interactions between mobile particles are taken into account they may be assumed to be spinless (spin-polarized) fermions. One could also imagine a scheme with bosons with  interactions tuned off by some microwave or optical Feshbach resonance \cite{Chin2010} (note that we already assume  using a magnetic field for a standard Feshbach tuning of different species interactions so this method cannot be used simultaneously to control light-light particle collisions).
The presented method allows us to obtain models with strong off-diagonal and weak diagonal disorder in a broad regime of relative
(off)-diagonal disorder amplitudes.
In a regime of weak off-diagonal disorder, an analytical expression for the Anderson localization length is in very good agreement with numerical simulations. Moreover, we indicate how DRDM with random hopping can work as a tunable band-pass filter for matter-waves. The momentum of escaped atoms reveal narrow peaks in the position of  the transparent mode in  momentum space while momentum distribution of the Anderson localized atoms reveals dips whose positions agree with former peaks.
We indicate the importance of the next-nearest-neighbor hopping for the  localization length of the resonant extended mode appearing within dual dimer random model for strong diagonal disorder. 
\section{Ackowlegments}
We are grateful to D. Delande, O. Dutta and K. Sacha for encouraging discussions and  for critical reading of the manuscript.
We acknowledge a support of the Polish National Science Centre via project DEC-2012/04/A/ST2/00088. AK acknowledges support in a form of a special scholarship of Marian Smoluchowski Scientific Consortium Matter Energy Future from KNOW funding.

 
 \section{Appendix A:Hamiltonian approach to Anderson localization length}
 
 To find the localization length starting from the map (\ref{eq:map}) \cite{flores1989,izrailev2001, izrailev1995}
   it is more convenient to express it in the action-angle variables $(r,\theta)$ using the transformation $x = r\sin(\theta)$, $p= r\cos(\theta)$:
 \begin{align}\label{action-angle-map}
 \left\{\begin{matrix}
 \sin(\theta_{i+1}) = D^{-1}_i (\sin(\theta_i - k) - A_i \sin(\theta_i)\sin (k))\\
 \cos(\theta_{i+1}) = D^{-1}_i(\cos(\theta_i -k) + A_i\sin(\theta_i)\cos(k))
 \end{matrix}\right.,
 \end{align}
 where
 \begin{align}
 D_i \equiv \frac{r_{i+1}}{r_i} = \sqrt{1+2 A_i \sin(2\theta_i) + A^2_i\sin^2(\theta_i)}.
 \end{align}
 We define Anderson localization length as
 \begin{eqnarray}
  \lambda^{-1}  =  \lim_{N\to\infty}\frac{1}{N} \sum_{i = 1}^{N} \ln D_i,
 \end{eqnarray}
 and after the expanding logarithm to the second order in $A_i$ we get
 \begin{equation}
  \lambda^{-1} = \frac{1}{8}\langle A^2_i\rangle+\frac{1}{2}\langle A_i \sin(2\theta_i)\rangle ,
 \end{equation}
where $\langle ... \rangle$ stands for averaging over $i$. 
In order to calculate the 'kick-angle' correlator $\langle A_i\sin (2\theta_i) \rangle$ we expand the map (\ref{action-angle-map}) to second order in $\theta_i$: 
\begin{equation}\label{theta_i_expanded}
 \theta_i = \theta_{i-1} - k + A_{i-1}\frac{\sin^2 (\theta_{i-1})}{\sin (k)}
\end{equation}
and express $\langle  A_i\sin (2\theta_i) \rangle$ in terms of the preceding $\langle  A_i\sin (2\theta_{i-1}) \rangle$ 'kick-angle' correlator. 
Let us introduce the correlation of the kick's strength $A_i$ with angle $\theta_i$ 
\begin{equation}
 a_n  = -\frac{2 i }{ \langle A_i^2 \rangle }e^{2 i k}\langle A_i e^{2 i \theta_{i-n}} \rangle,\\
\end{equation}
where $\langle A_i^2 \rangle$ is variance of the $A_i$.
Multiple applications of eq. (\ref{theta_i_expanded}) to $a_n$ give us the recursive relation
\begin{equation}\label{a_0}
 a_{n-1} = e^{-2i k} a_n + q_n,
\end{equation}
where $q_n  =  \langle A_i A_{i-n}\rangle / \langle A_i^2 \rangle$ is the autocorrelation of $A_i$. 
From the definition of $a_n$ we can notice that
\begin{equation}
 \langle A_i \sin(2\theta_i) \rangle = \Re\left( \frac{\langle A_i^2 \rangle}{2} e^{-2 i k}a_0 \right),
\end{equation}
where from (\ref{a_0}) we obtain $a_0 = \sum_{n=1}^{\infty} q_ne^{-2ik (n-1)}$. \\

The inverse Anderson localization length $\lambda^{-1}$ {takes the form \cite{izrailev2001}}
\begin{equation}
 \lambda^{-1} = \frac{\langle A_i^2 \rangle}{8}\left(1 + 2\sum_{n=1}^{\infty}q_n(k)\cos(2 k n)\right).\\
\end{equation}
In the specific case of DRDM , the variance of $A_i$ and the autocorrelation of $A_i$ read, respectively
\begin{eqnarray}
\langle A^2_i\rangle &=&\nonumber\frac{\rho}{(1+\rho)^2}\frac{(V_0+2\gamma(2-\gamma)\cos (k))^2}{\sin^2(k)},\\
 q_n & = & (-1)^n \rho^n
\end{eqnarray}

and the approximate expression for the inverse localization length takes the form (\ref{lambda}).

\section{Appendix B:Anderson localization length for Hamiltonian with long-range random hopping}
In this section we describe a method of calculating localization length in a one dimensional system with long-range tunneling. The idea is to transform the problem from a one dimensional to a two dimensional stripe. It could be achieved with a simple winding of a string, see~Fig.~\ref{stripe}.

\subsection{Mapping to a two--dimensional stripe}

To start let us consider a one dimensional gas on a string of length $\Nt$. We allow long-range tunneling up to the $M$-th neighboring site. This geometry is equivalent to a two dimensional stripe of length $N=\Nt/M$ and width $M$. The evolution of particles in such a stripe is characterized by a Hamiltonian
\be
H(\Nt) = \sum_{n=1}^{N} H_n + \sum_{n=1}^{N-1} \Big( V_n+h.c. \Big),
\ee
where
\bea
H_n&=& \sum_{m=1}^{M} \epsilon^{(n)}_m |n,m \rangle \langle n,m| \nonumber \\
&-&\sum_{m=1}^{M-1} \Big(J^{(n,n)}_{m,m+1}|n,m\rangle \langle n,m+1| +h.c. \Big)
\eea
is a standard tight binding Hamiltonian for the $n$-th slice and
\be
V_n=-\sum_{m=1}^M\sum_{m'=1}^{M'}J^{(n,n+1)}_{m,m'}|n,m\rangle \langle n+1,m'|
\ee
describes particles hopping onto the $n$-th slice. $J^{(n,n')}_{m,m'}$ denotes tunneling between $|n,m\rangle$ and $|n',m'\rangle$. Suppose now that we add one extra slice to the system. A new Hamiltonian is straightforward
\be
H(N+1)=H(N)+H_{N+1}+V_{N}+V_{N}^{\dagger}.
\ee
For later convenience, let us define operators
\bea
H^0(N+1) &\equiv& H(N)+H_{N+1} \label{H_0} \\
V(N) &\equiv&  V_{N}+V_{N}^{\dagger},
\eea
and finally the Hamiltonian reads
\be
H(N+1) = H^0(N+1) + V(N) .
\ee

\subsection{A recursive equation for a Green's function} 

Our goal is to find the Green's function $G^+_{E}(N+1)$ for the Hamiltonian $H(N+1)$. By definition, a Green's function can be obtained from a resolvent of a Hamiltonian
\be
G^+_{E}(N+1) = \lim_{\epsilon \rightarrow 0+} G_{E+i\epsilon}(N+1), 
\ee
where a resolvent satisfies 
\be
\left(z-H(N+1)\right)G_z(N+1) =1. \label{resolvent_def}
\ee

Let's start with an equation for a resolvent of $H(N+1)$:
\be
G_z(N+1) = G_z^0(N+1) + G_z^0(N+1)V(N)G_z(N+1),
\label{resolvent_eq}
\ee
with $G^0_z$ being a resolvent of $H^0$. The resolvent equation~(\ref{resolvent_eq}) can be derived from a simple operator identity 
\be
\frac{1}{A} = \frac{1}{B} + \frac{1}{B}(A-B)\frac{1}{A}  
\ee
with a substitution
\be
A=z-H(N+1), \quad B=z-H^0(N+1).
\ee

Since we are interested in transport properties of the system, we need to know the Green's function matrix elements between the first and the last slice only. Therefore,  we want to obtain $\langle 1 | G_z(N+1) | N+1 \rangle $. To simplify the equations  let's introduce a notation:
\be
G_{n,m} \equiv \langle n | G_z(m) | m \rangle  .
\ee
From equation (\ref{resolvent_eq}) we get
\be
G_{1,N+1}=G_{1,N} V_N \,G_{N+1,N+1},
\label{resolvent_1}
\ee
where we used an observation that 
\be
G^0_z(N+1) = G_z(N) +\frac{1}{z-H_{N+1}},
\ee
which stems from the fact that $H(N)$ and $H_{N+1}$ act in orthogonal subspaces of the Hilbert space, see (\ref{H_0}). 

In order to get a recursive equation for $G_{1,N}$, we need to calculate $G_{N+1,N+1}$ matrix. It can be obtained from equation (\ref{resolvent_def}) by multiplying it from the right side by a projection $P=|N+1 \rangle \langle N+1|$, and by $P$ (or $Q=1-P$)  from the left side:
\bea
P(z-H(N+1))(P+Q) G_z(N+1)P &=& P, \nonumber \\
Q(z-H(N+1))(P+Q) G_z(N+1)P &=& 0 .
\label{proj_green_equations}
\eea
Solving a set of equations (\ref{proj_green_equations}) one gets
\be
G_{N+1,N+1}=\frac{1}{z-H_{N+1}-V_N^\dagger G_{N,N} V_N} .
\label{resolvent_2}
\ee
Combining equations (\ref{resolvent_1}) and (\ref{resolvent_2})
\be
G_{1,N+1} \left( z-H_N+1 -V_N^\dagger G_{N,N} V_N \right) = G_{1,N} V_N
\ee
and, again from (\ref{resolvent_1}), extracting $G_{N,N}$ we  finally obtain
a recursive equation:
\be
A_{N+2} = (E-H_{N+1})V_N^{-1}A_{N+1} -V_N^\dagger V_{N-1}^{-1} A_{N},
\label{recursive_equation}
\ee
with $A_N=G^{-1}_{1,N-2}$ . The initial values of $A_N$ are not relevant, but it is convenient to choose them as
\be 
A_0=0,  \quad A_1=V_0.
\ee
In equation (\ref{recursive_equation}) there is no singularity, hence we could replace $z$ with $E$. From this equation we can calculate $G_{1,N}$, which is connected with the localization length of the system.

\subsection{Calculation of the Anderson localization length} 

The Anderson localization length $\lambda_M$ in a stripe of width $M$ is defined as 
\be
\frac{2}{\lambda_M} = - \lim_{n\rightarrow\infty} \frac{1}{n} \ln \mbox{Tr} |G_{1,n}|^2. 
\ee
Since our system is in fact one dimensional the localization length should not depend on the length of the stripe. When a particle reached the $n$-th slice it covered $\tilde{n}= M \cdot n$ lattice sites. Hence, the localization length $\lambda$ in one dimension equals $\lambda= M \lambda_M.$  

The localization length can be obtained from the recursive equation  (\ref{recursive_equation}). We solve the equation iteratively.  The problem with this equation is that the elements of $A_n$ grow exponentially for large $n$, so it requires some regularization. Therefore, in each step we multiply the both sides of the equation by some matrix $R_n$. Starting from~$n=1$ and defining $A^{(1)}_k \equiv A_k R_1$ 
\bea
A_3 &=& (E-H_2)V_1^{-1}A_{2} -V_1^\dagger V_0^{-1} A_1 \; \Big| \times R_1 \nonumber \\
A_3^{(1)} &=& (E-H_2)V_1^{-1}A_{2}^{(1)} -V_1^\dagger V_0^{-1} A_1^{(1)}.
\eea
To avoid the exponential growth we put  $R_1= A_2^{-1}$ and
\be
A_3^{(1)}=A_3 A_2^{-1}, \quad A_2^{(1)}=1, \quad A_1^{(1)} = A_1 A_2^{-1}.
\ee
We repeat this procedure in every iteration so that 
$ A_k^{(n)} = A_k^{(n-1)} R_n$, with $R_n=\left[ A^{(n-1)}_{n+1} \right]^{-1}$, satisfies 
\be
A_{n+2}^{(n)} = (E-H_{n+1})V_n^{-1} -V_n^\dagger V_{n-1}^{-1} A_n^{(n)}.
\ee
Let us also define a matrix
\be
B^{(n)} = B^{(n-1)} R_n /b_n, \quad b_n = \| B^{(n-1)} R_n \|,
\ee
where $\| \cdot  \| = \sqrt{ \mbox{Tr}|\cdot|^2}$ is a matrix norm.  $B^{(n)}$ matrix turns out to be very useful, because
\bea
b_n &=& \| B^{(n-1)} R_n \|= \frac{1}{b_{n-1}}\left\| B^{(n-2)} R_{n-1} R_n\right\| = \nonumber \\
&=& \frac{1}{b_{n-1}} \left\| B^{(n-2)} \left[ A_{n+1}^{(n-2)}\right]^{-1} \right\| = \nonumber \\
&=& \frac{1}{b_{n-1}b_{n-2}} \left\| B^{(n-3)} \left[ A_{n+1}^{(n-3)}\right]^{-1} \right\| = \ldots\, .
\eea
Continuing  this, we notice that
\bea
\|A_{n+1}^{-1}\| &=& b_1 b_2 \ldots b_n, \\
 \ln \mbox{Tr} |G_{1,n}|^2 &=& 2 \left( \ln b_{n+1} +  \ldots + \ln b_1 \right).
\eea
The Anderson localization length can be expressed as: 
\be
\lambda= -M \lim_{n\rightarrow \infty} \frac{n}{c_{n+1}}, \quad c_{n+1}=c_n +\ln b_{n+1}. 
\ee
Iterations should be continued until $\lambda$ converges within a desired precision. 
 
\end{document}